\documentclass[twocolumn,prl,aps,showpacs]{revtex4}

\usepackage{psfrag,graphicx}
\usepackage{dcolumn}
\usepackage{amsmath,amssymb}
\usepackage{bm}
\usepackage[dvips]{color}

\definecolor{mygrey}{gray}{0.35}
\definecolor{mygreen}{rgb}{0.85,1,0.9}
\definecolor{myzard}{cmyk}{0,0,0.05,0}
\definecolor{mywhite}{rgb}{1,1,1}
\definecolor{myred}{rgb}{1,0,0}

 \def\ee{\mathord{\rm e}}
 
 \def\ii{\mathord{\rm i}}

\def\half{\textstyle\frac{1}{2}}

\renewcommand{\ii}{{\rm i}}
\renewcommand{\ee}{{\rm e}}

 \newcommand{\ket}[1]{|#1\rangle}
 \newcommand{\bra}[1]{\langle #1|}

\begin{document}

\title[Short Title]{Exact Mapping of the 2+1 Dirac Oscillator onto the
Jaynes-Cummings Model: Ion-Trap Experimental Proposal}

\author{A. Bermudez$^1$, M.~A. Martin-Delgado$^1$ and E. Solano$^{2,3}$}

\affiliation{ $^1$Departamento de F\'{\i}sica Te\'orica I,
Universidad Complutense, 28040 Madrid, Spain \\$^2$ Physics
Department, ASC, and CeNS, Ludwig-Maximilians-Universit\"at,
Theresienstrasse 37, 80333 Munich, Germany \\$^3$Secci\'on
F\'isica, Departamento de Ciencias, Pontificia Universidad
Cat\'olica del Per\'u, Apartado Postal 1761, Lima, Peru }

\begin{abstract}
We study the dynamics of the 2+1 Dirac oscillator exactly and find
spin oscillations due to a {\it Zitterbewegung} of purely
relativistic origin. We find an exact mapping of this
quantum-relativistic system onto a Jaynes-Cummings model, describing the interaction of a two-level
atom with a quantized single-mode field. This equivalence allows
us to map a series of  quantum optical phenomena onto the
relativistic oscillator, and viceversa. We make a realistic
experimental proposal, at reach with current technology, for
studying the equivalence of both models using a single trapped
ion.
\end{abstract}


\maketitle

Current technology has allowed the implementation of the
paradigmatic nonrelativistic quantum harmonic oscillator in a
single trapped ion~\cite{wineland_review}, one of the most
fundamental toy models in any quantum mechanical textbook.
However, its relativistic version, the so-called Dirac oscillator
\cite{imc67,moshinsky}, remains still far from any possible
experimental consideration for different fundamental and technical
reasons. We will show here that available experimental tools may
allow the implementation of the relativistic Dirac oscillator in a
single nonrelativistic trapped ion.

The Dirac oscillator was introduced as an instance of a
relativistic wave equation such that its nonrelativistic limit
leads to the well-known Schr\"odinger equation for the harmonic
oscillator. This is achieved by introducing the following coupling
in the Dirac equation
\begin{equation}
\label{ec_dirac_oscillator_3}
 \ii \hbar\frac{\partial |\Psi\rangle}{\partial t}=\left[\sum_{j=1}^3c\alpha_j\left(p^j-\ii m\beta\omega r^j\right)+\beta
mc^2\right]|\Psi\rangle,
\end{equation}
where $| \Psi \rangle$ is the Dirac 4-component bispinor
corresponding to a quantum relativistic spin-$\half$ particle,
like the electron, $c$ is the speed of light, $m$ is the particle
rest mass, and $\alpha_j$, $\beta$, are the Dirac matrices in the
standard representation. The interacting Hamiltonian is linear in
both momentum $p^j$ and position $r^j$, $j = x,y, z$, and $\omega$ turns out to be the harmonic oscillator
frequency. Remark that when $\omega = 0$  we recover the standard
Dirac equation~\cite{greiner}. The Dirac oscillator looks like a
particular gauge transformation $\mathbf p \rightarrow \mathbf p -
\frac{e}{c} \mathbf A$ that is linear in position, but the
presence of the $\ii $ and the $\beta$ matrix makes a crucial
difference. Demanding the correct energy-momentum relation for a
relativistic free particle $E=\sqrt{p^2c^2+m^2c^4}$, these
matrices are $4\times 4$ dimensional and must obey a Clifford
algebra given by the anticommutation relations
\begin{equation}
\label{ant_dirac}
\begin{array}{c}
  \alpha_j\alpha_k+\alpha_k\alpha_j=2\delta_{jk}, \\
  \alpha_j\beta+\beta\alpha_j=0. \\
\end{array}
\end{equation}

There has been a growing interest in simulating quantum
relativistic effects in other physical systems, such as black hole
evaporation in Bose-Einstein condensates~\cite{garay} and the
Unruh effect in an ion chain~\cite{alsing}. Another astonishing
relativistic prediction is the {\it
Zitterbewegung}~\cite{greiner}, a helicoidal motion realized by
the average position of a relativistic fermion, which has been
discussed in the context of condensed matter
systems~\cite{schliemann} and the free-particle Dirac equation in
a single ion~\cite{lamata}.

Here,  we shall be concerned with the Dirac
oscillator model in 2+1 dimensions, since it is in this setting
where we can establish a precise equivalence with the
Jaynes-Cummings (JC) model~\cite{jaynes_cummings}. In two spatial dimensions, the
solution to the Clifford algebra \eqref{ant_dirac} is given by the
$2\times 2$ Pauli matrices:
$\alpha_x=\sigma_x,\alpha_y=\sigma_y,\beta=\sigma_z$. In this case, $|\Psi\rangle$ can be described by a 2-component spinor
which mixes spin up and down
components with positive and negative energies. In particular,
the Dirac oscillator model now takes the form
\begin{equation}
\label{ec_dirac_oscillator}
 \ii \hbar\frac{\partial |\Psi\rangle}{\partial t}=\left[\sum_{j=1}^2c\sigma_{j}\left(p^j-\ii m\sigma_z\omega
r^j\right)+\sigma_z mc^2\right]|\Psi\rangle.
\end{equation}
In this paper, we shall provide the complete (eigenstates and energies) and
exact solution of the 2D Dirac oscillator in order to study its
relativistic dynamics, where certain collapses and revivals in the
spin degree of freedom appear as a consequence of {\it Zitterbewegung}.
In addition, we derive an exact mapping of the 2+1 Dirac oscillator onto the JC model, an archetypical quantum optical system. Furthermore, we propose the simulation of this relativistic dynamics in a single trapped ion, a physical setup possessing outstanding coherence features.

Considering the spinor $|\Psi\rangle:= [ | \psi_1
\rangle ,| \psi_2 \rangle ]^t$,
equation~\eqref{ec_dirac_oscillator} becomes a set of coupled
equations
\begin{equation}
\label{ec_do_components}
\begin{array}{c}
  (E-mc^2)| \psi_1 \rangle = c \left[(p_x+\ii m\omega x)-\ii(p_y+\ii m\omega y)\right] |\psi_2 \rangle ,\\
  (E+mc^2) \ket{\psi_2} =c\left[(p_x-\ii m\omega x)+\ii(p_y-\ii m\omega y)\right] \ket{\psi_1}.
\end{array}
\end{equation}
In order to find the solutions, it is convenient to  introduce the
following chiral creation and annihilation operators
\begin{equation}
\label{circular_operators}
\begin{array}{c}
  a_r:=\frac{1}{\sqrt{2}}(a_x - \ii a_y),\hspace{2ex}a_r^\dagger:=\frac{1}{\sqrt{2}}(a_x^\dagger + \ii a_y^\dagger) , \\
  a_l:=\frac{1}{\sqrt{2}}(a_x + \ii a_y), \hspace{2ex}a_l^\dagger:=\frac{1}{\sqrt{2}}(a_x^\dagger - \ii a_y^\dagger) , \\
\end{array}
\end{equation}
where $a_x, a_x^\dagger, a_y, a_y^\dagger$, are the usual
annihilation and creation operators of the harmonic oscillator
$a^{\dagger}_i=\frac{1}{\sqrt{2}}\left(\frac{1}{\Delta}r^i - \ii
\frac{\Delta}{\hbar}p^i\right)$, and $\Delta=\sqrt{\hbar/m\omega}$
represents the ground state oscillator width. The orbital angular
momentum may also be expressed as
\begin{equation}
\label{ang_momentum}
 L_z=\hbar(a_r^\dagger a_r-a_l^\dagger a_l),
\end{equation}
which leads to a physical interpretation of $a_{r}^\dagger$ and
$a_{l}^\dagger$. These operators create a right or left quantum of
angular momentum, respectively, and are known hence as circular
creation-annihilation operators. Equations~\eqref{ec_do_components} can be rewritten in the language of
these circular operators
\begin{equation}
\label{circular_DO}
\begin{array}{c}
  |\psi_1\rangle=\hspace{2ex}\ii\frac{2 m c^2 \sqrt{\xi}}{E-mc^2}\ a_l^\dagger \ |\psi_2\rangle ,\\
  |\psi_2\rangle=-\ii\frac{2 m c^2 \sqrt{\xi}}{E+mc^2}\ a_l \  |\psi_1\rangle,\\
\end{array}
\end{equation}
where $\xi := \hbar\omega/mc^2$ controls the nonrelativistic
limit. In order to find the energy spectrum we shall solve the
associated Klein-Gordon equation, which can be derived from
Eqs.~\eqref{circular_DO} as follows
\begin{equation}
\begin{split}
  &(E^2-m^2c^4)|\psi_1\rangle=4m^2c^4\xi\ a_l^\dagger a_l \ |\psi_1\rangle ,\\
  &(E^2-m^2c^4)|\psi_2\rangle= 4m^2c^4\xi\ (1+a_l^\dagger a_l)  |\psi_2\rangle. \\
\end{split}
\end{equation}
These equations can be simultaneously diagonalized writing the
spinor in terms of the left chiral quanta basis
\begin{equation}
|n_l\rangle=\frac{1}{\sqrt{n_l!}}\left(a_l^\dagger\right)^{n_l}|\text{vac}\rangle,
\end{equation}
where $n_l = 0,1, ...$ The energies can be
expressed as
\begin{equation}
\begin{split}
  &(E_{n_l}^2-m^2c^4)|n_l\rangle=4m^2c^4\xi n_l \ |n_l\rangle ,\\
  &(E_{n_l'}^2-m^2c^4)|n_l'\rangle= 4m^2c^4\xi\ (1+n_l')  |n_l'\rangle. \\
\end{split}
\end{equation}
Since both components $\ket{\psi_1}$ and $\ket{\psi_2}$ belong to
the same solution, the energies must be the same $
E_{n_l'}=E_{n_l}$. This physical requirement sets up a constraint
on the quantum numbers $n_l =: n_l'+1$. Note that, following
\eqref{ang_momentum}, the state $\ket{ n_l}$ corresponds to a
negative angular momentum. The energy spectrum can be described as
follows
\begin{equation}
E=\pm E_{n_l}=\pm mc^2\sqrt{1+4\xi n_l}.
\end{equation}
To find the corresponding eigenstates, we go back
to Eq.~\eqref{circular_DO}, and after normalization we arrive at
the expression for the positive and negative energy
eigenstates
\begin{equation}
| \pm E_{n_l} \rangle=\left[
\begin{array}{c}
  \sqrt{\frac{E_{n_l}\pm mc^2}{2E_{n_l}}}|n_l\rangle \vspace{1ex}\\
  \mp \ii\sqrt{\frac{E_{n_l}\mp mc^2}{2E_{n_l}}}|n_l-1\rangle \\
\end{array}
\right],
\end{equation}
where the quantum number is now restricted to $n_l=1,2, ...$ In
this way, we have solved the two-dimensional Dirac oscillator
describing the energy spectrum and the eigenstates in terms of
circular quanta. The distinction between Dirac and Klein-Gordon
eigenstates is an important point in order to understand the
dynamics of the 2+1 Dirac oscillator and its realization in an ion
trap.

The eigenstates of the 2D Dirac oscillator can be expressed more
transparently in terms of 2-component Pauli spinors
$\ket{\chi_{\uparrow}}$ and $\ket{\chi_{\downarrow}}$
\begin{equation}
\begin{array}{c}
  | +E_{n_l} \rangle = \alpha_{n_l} |n_l\rangle \ket{\chi_\uparrow} - \ii \beta_{n_l}|n_l-1\rangle \ket{\chi_\downarrow}, \\
  | -E_{n_l} \rangle=\beta_{n_l}|n_l\rangle \ket{\chi_\uparrow} + \ii \alpha_{n_l}|n_l-1\rangle \ket{\chi_\downarrow}, \\
\end{array}
\end{equation}
where $\alpha_{n_l}:=\sqrt{\frac{E_{n_l}+mc^2}{2E_{n_l}}}$ and
$\beta_{n_l}:=\sqrt{\frac{E_{n_l}-mc^2}{2E_{n_l}}}$ are real. From
this expression we observe that the energy eigenstates present
entanglement between the orbital and spin degrees of freedom. This
property is extremely important since the following initial state
\begin{equation}
|\Psi(0)\rangle:=|n_l-1\rangle \ket{\chi_\downarrow} = \ii
\beta_{n_l}| +E_{n_l} \rangle - \ii \alpha_{n_l} |-E_{n_l}\rangle
\end{equation}
superposes states with positive and negative energies, and this is
the fundamental ingredient that leads to {\it Zitterbewegung} in
relativistic quantum dynamics. This phenomenon, due to the
interference of positive and negative energies, has never been
observed experimentally. The reason is that the
amplitude of these rapid oscillations lies below the Compton
wavelength, where pair creation is allowed, and the one-particle
interpretation falls down.

Now, the evolution of this initial state can be expressed in
the energy basis as
\begin{equation}
|\Psi(t)\rangle=\ii \beta_{n_l}|+E_{n_l}\rangle
\ee^{-\ii\omega_{n_l}t}-\ii \alpha_{n_l}|-E_{n_l}\rangle
\ee^{\ii\omega_{n_l}t},
\end{equation}
where
\begin{equation}
\label{ZBfrequency} \omega_{n_l} := \frac{E_{n_l}}{\hbar} =
\frac{m c^2}{\hbar} \sqrt{1+4\xi n_l}
\end{equation}
describes the frequency of oscillations. Writing this evolved
state in the language of Pauli spinors,
\begin{equation}
\label{state_dynamics}
\begin{split}
|\Psi(t)\rangle = & \left(\cos\omega_{n_l}t+\frac{\ii}{\sqrt{1+4\xi
n_l}}\sin\omega_{n_l}t\right)|n_l-1\rangle \ket{\chi_\downarrow} \\
& + \left(\sqrt{\frac{4\xi n_l}{1+4\xi
n_l}}\sin\omega_{n_l}t\right) |n_l\rangle \ket{\chi_\uparrow},
\end{split}
\end{equation}
we observe an oscillatory dynamics between $ |n_l \rangle
\ket{\chi_\uparrow}$ and  $|n_l-1\rangle \ket{\chi_\downarrow}$.
The initial state, $|n_l-1\rangle \ket{\chi_\downarrow}$, which
has spin-down and $n_l-1$ quanta of left orbital angular momentum,
evolves exchanging a quantum of angular momentum from the spin to
the orbital motion.

The dynamics described in \eqref{state_dynamics} is completely
similar to the atomic Rabi oscillations occurring in the
Jaynes-Cummings model, though arising from a completely different
reason. Whereas the Rabi oscillations in the Jaynes-Cummings model
are caused by the interaction of a quantized electromagnetic field
with a two-level atom, the relativistic oscillations are caused by
the interference of positive and negative energy states and
therefore constitute a clear signature of {\it
Zitterbewegung}~\cite{greiner}.

To clarify this issue further, we calculate the time evolution of the following physical
observables, that catch the full essence of the system dynamics,
\begin{equation}
\label{zitter_dynamics}
\begin{array}{l}
  \langle L_z\rangle_{t}=-(n_l-1)\hbar-\frac{4\xi n_l}{1+4\xi
n_l}\hbar\sin^2\omega_{n_l}t, \\
    \langle S_z\rangle_t=\hspace{4ex} -\frac{\hbar}{2}\hspace{1.5ex} \hspace{2ex}+\frac{4\xi n_l}{1+4\xi
n_l}\hbar\sin^2\omega_{n_l}t ,  \\
\langle J_z\rangle_t=\hspace{1ex}\hbar(\half - n_l),
\end{array}
\end{equation}
where $J_z=L_z+S_z$ stands for the $z$-component of the total
angular momentum. The latter relations describe a certain
oscillation in the spin and orbital angular momentum, while the
total angular momentum is conserved due to the existent invariance
under rotations around the z-axis. It is important to highlight
that these oscillations have a pure relativistic nature. In the
nonrelativistic limit $\xi\ll1$, these oscillations become
vanishingly small
\begin{equation}
\begin{array}{l}
  \langle L_z\rangle_t=-(n_l-1)\hbar-4\xi n_l\hbar\sin^2\Omega_{n_l}t+\mathcal{O}(\xi^2), \\
    \langle S_z\rangle_t=\hspace{4ex} -\frac{\hbar}{2} \hspace{3.5ex}+4\xi n_l\hbar\sin^2\Omega_{n_l}t+\mathcal{O}(\xi^2), \\
\end{array}
\end{equation}
where $\Omega_{n_l} := mc^2(1+2\xi n_l)/\hbar$ stands for the
oscillation frequency in the nonrelativistic limit. In this limit
the negative energy components are negligible and therefore the
{\it Zitterbewung} disappears.

The results discussed so far allow a precise mapping between two seemingly unrelated models: the Jaynes-Cummings model of Quantum Optics and the 2D Dirac
oscillator. Starting from
Eq.~\eqref{circular_DO}, we may write the Dirac oscillator
Hamiltonian as
\begin{equation}
\label{JC_DO}
\begin{split}
H&=2 \ii mc^2 \sqrt{\xi}\left(a_l^\dagger|\psi_2\rangle\langle\psi_1|-a_l|\psi_1\rangle\langle\psi_2|\right)+mc^2\sigma_z \\
&=\hbar(g\sigma^-a_l^\dagger+g^*\sigma^+a_l)+mc^2\sigma_z,
\end{split}
\end{equation}
where $\sigma^+$, $\sigma^-$, are the spin raising and lowering
operators, and $g := 2 \ii mc^2 \sqrt{\xi} / \hbar$ is the coupling strength
between orbital and spin degrees of freedom. In Quantum Optics,
this Hamiltonian describes a Jaynes-Cummings interaction, that
has been studied in cavity QED and trapped ions \cite{haroche_review,wineland_review}, among others. Within
this novel perspective, the electron spin can be associated with a
two-level atom, and the orbital circular quanta with the ion quanta
of vibration, i.e., phonons. As we will see below, the central result of Eq.~\eqref{JC_DO} allows both physical systems, the JC model and the 2D Dirac oscillator, to exchange a wide range of important applications.

We will show now how to implement the dynamics of
Eq.~\eqref{ec_dirac_oscillator} in a single ion inside a Paul
trap, which was shown to follow the dynamics of Eq.~\eqref{JC_DO}.
The Dirac spinor will be described by two metastable internal
states, $\ket{g}$ and $\ket{e}$, as follows
\begin{equation}
\ket{\Psi}:=\ket{\psi_1}\ket{e}+\ket{\psi_2}\ket{g}
\end{equation}
while the circular angular momentum modes will be represented by
two ionic vibrational modes, $a_x$ and $a_y$. Current technology
allows an overwhelming coherent control of ionic internal and
external degrees of freedom~\cite{wineland_review}. There, three
paradigmatic interactions, the carrier, red-, and blue-sideband
excitations, can be implemented at will, independently or
simultaneously~\cite{comment1}. For example, using appropriately
tuned lasers, it is possible to produce the following interactions
\begin{equation}
\label{JC_ham}
\begin{array}{l}
H_i^{\rm JC}\hspace{1.5ex}=\hbar\eta_i\tilde{\Omega}_i\left[\sigma^+a_i \ee^{\ii \phi}+\sigma^-a_i^{\dagger} \ee^{-\ii \phi}\right]+\hbar \delta_i \sigma_z, \\
H_i^{\rm AJC}=\hbar\eta_i\tilde{\Omega}_i\left[\sigma^+a^\dagger_i
\ee^{\ii \varphi}+\sigma^-a_i \ee^{-\ii \varphi}\right],
\end{array}
\end{equation}
where $\{a_i,a^\dagger_i\}$, with $i=x,y$, are the phonon
annihilation and creation operators in directions $x$ and $y$,
$\nu_i$ are the natural trap frequencies,
$\eta_i:=k_i\sqrt{\hbar/2M\nu_i}$ are the associated Lamb-Dicke
parameters depending on the ion mass $M$ and the wave vector
$\bold{k}$, $\delta_i$ and $\tilde{\Omega}_i$ are the excitation
coupling strengths and $\phi,\varphi$, the red and blue sideband
phases. Remark that the term $\hbar \delta_i \sigma_z$, in
$H_i^{\rm JC}$ of Eq.~\eqref{JC_ham}, stems from a detuned JC
excitation.

A suitable combination of the above introduced excitations~\eqref{JC_ham}, with proper couplings and relative phases, can reproduce the following Hamiltonian
\begin{equation}
\label{do_iontrap}
H = c\left[\sigma^{ge}_xp_x+\sigma^{ge}_yp_y\right] + m \omega
c \left[\sigma_x^{ge}y-\sigma^{ge}_yx\right] + mc^2 \sigma^{ge}_z
\end{equation}
with $ \sigma^{ge}_x := \ket{g} \bra{e} + \ket{g} \bra{e}$, $ \sigma^{ge}_y := -\ii ( \ket{e} \bra{g} - \ket{e} \bra{g})$, $ \sigma^{ge}_z := \ket{e} \bra{e} - \ket{g} \bra{g}$, and the following parameter correspondence
\begin{equation}
\label{exper_parameter}
\begin{array}{l}
  c=\sqrt{2}\eta\tilde{\Omega} \tilde\Delta, \\
  mc^2=\hbar \delta, \\
  m\omega c=\hbar\sqrt{2}\eta\tilde{\Omega} \tilde\Delta^{-1} , \\
\end{array}
\end{equation}
where
$\tilde\Delta := \tilde\Delta_i$ is the width of the motional ground state, $\tilde{\Omega}:=\tilde{\Omega}_i,\eta:=\eta_i
, \forall i=x,y$. The remarkable equivalence of the Dirac
oscillator Hamiltonian~\eqref{ec_dirac_oscillator} and the interaction in Eq.~\eqref{do_iontrap} shows that it is
possible to reproduce the 2D Dirac oscillator, with all its
quantum relativistic effects, in a controllable quantum system as
a single trapped ion.

For the sake of illustration, note that the effective terms appearing in Eq.~\eqref{do_iontrap} can be achieved by suitable linear combinations of $H_i^{\rm JC}$ and $H_i^{\rm AJC}$ in \eqref{JC_ham},
\begin{equation}
\label{dirac_momentum}
\begin{array}{l}
  i=x,\delta_x = \delta, \phi = \frac{3\pi}{2},\varphi = \frac{\pi}{2}\rightarrow  \sqrt{2}\hbar\eta \tilde{\Omega} \tilde{ \Delta} \sigma^{ge}_xp_x + \hbar \delta \sigma^{ge}_z ,\\
  i=y, \delta_y = 0, \phi = 0 \hspace{1.7ex}, \varphi = \pi \rightarrow \sqrt{2} \hbar \eta \tilde{\Omega} \tilde{ \Delta} \sigma^{ge}_y p_y, \\
  i=x,\delta_x = 0, \phi = \frac{\pi}{2} , \hspace{1ex}\varphi = \frac{\pi}{2}\rightarrow \sqrt{2} \hbar \eta \tilde{\Omega} \tilde{ \Delta} ^{-1}\sigma^{ge}_y x,\\
i=y ,\delta_y=0, \phi = 0 \hspace{1ex}, \hspace{1ex}\varphi = 0 \rightarrow  \sqrt{2} \hbar \eta \tilde{\Omega} \tilde{ \Delta }^{-1} \sigma^{ge}_x y. \\
\end{array}
\end{equation}
Note that in the trapped ion picture, the important parameter $\xi = 2 (\eta \tilde\Omega / \delta)^2$ can take on all positive values, assuming available experimental parameters: $\eta \sim 0.1$, $\tilde\Omega \sim 0-10^6 {\rm Hz}$, and $\delta \sim 0-10^6 {\rm Hz}$~\cite{wineland_review}. The ability to experimentally tune these
parameters will allow the experimenter to study otherwise
inaccessible physical regimes that entail relativistic and nonrelativistic phenomena. For example, the {\it Zitterbewegung} is encoded in the spin degree of
freedom, and we can associate Rabi oscillations to the
interference of positive and negative energy solutions. Setting
the initial state $\ket{0}\ket{\chi_{\downarrow}} \leftrightarrow \ket{0}\ket{g}$, the internal degree of freedom
evolves according to Eq.~\eqref{zitter_dynamics}
\begin{equation}
\label{sinnombre}
 \langle S_z\rangle_t= -\frac{\hbar}{2}+\frac{4\xi}{1+4\xi
}\hbar\sin^2\omega_{1}t ,
\end{equation}
where $\omega_{1}= \delta \sqrt{1+4\xi}$, see Eq.~\eqref{ZBfrequency}, stands for the frequency of
the {\it Zitterbewegung} oscillations and can take on a wide variety of measurable values.

In order to simulate this dynamics in an ion-trap tabletop experiment, the
ion must be cooled down to its vibrational ground state $| 0 \rangle$, with a current efficiency above $99\%$~\cite{wineland_review}. To estimate the observable \eqref{sinnombre}, one can make use of the powerful tool called electron shelving, where
\begin{equation}
\langle S_z \rangle_t =\frac{\hbar}{2}\left[2P_e(t)-1\right]
\end{equation}
can be obtained through the measurement of the probability of obtaining the ionic excited state $P_e(t)$ with extraordinary precision.

Another fundamental result of the JC model which can be mapped
straightforward to the Dirac oscillator is the existence of
collapses and revivals in the atomic population, which is claimed
to be a direct evidence of the quantization of the electromagnetic
field. To produce this effect an initial state $| z \rangle | g \rangle$ is required, where $| z \rangle$ is an initial circular coherent state,
\begin{equation}
\label{circular_coherent}
|\Psi(0)\rangle=e^{-|z|^2/2}\sum_{n_l=0}^\infty
\frac{z^{n_l}}{\sqrt{n_l !}}|n_l \rangle | g \rangle,
\end{equation}
with $z\in \mathbb{C}$. After an interaction time $t$,
\begin{equation}
\label{ZBsz}
\langle S_z\rangle_t = -\frac{\hbar}{2}+\hbar\sum_{n_l=0}^\infty
\frac{4\xi(n_l+1)|z|^{2n_l}e^{-|z|^2}}{[1+4\xi
(n_l+1)]n_l!}\sin^2(\omega_{n_l+1}t).
\end{equation}
This expression can be understood as an interference effect of
terms with different frequencies
$\omega_{n_l+1}$ leading to
collapses and revivals. A novel feature of the Dirac oscillator
is the appearance of these collapses and revivals in the orbital
circular motion of the particle, reflected in
\begin{equation}
\label{sinnombre2}
\langle L_z\rangle_t=-\hbar|z|^2-\hbar\sum_{n_l=0}^\infty
\frac{4\xi(n_l+1)|z|^{2n_l}e^{-|z|^2}}{[1+4\xi
(n_l+1)]n_l!}\sin^2(\omega_{n_l+1}t).
\end{equation}
The generation of an initial circular coherent state will require two sequential applications of the technique described in Ref.~\cite{wineland_review} on an initial motional ground state. These two operations should be applied with a relative phase such that $D_l(z)=D_x(z) D_y(-\ii z)$, where $D_j(z)=\ee^{z a_j^{\dagger} - z^* a_j}$, $j=x,y$. The observable of Eq.~\eqref{ZBsz} can be measured via a similar electron-shelving technique, while the observable of Eq.~\eqref{sinnombre2} can be measured via the mapping of the collective motional state onto the internal degree of freedom~\cite{wineland_review}.

It is worth mentioning that the chiral partner of the 2D Dirac
oscillator Hamiltonian~\eqref{ec_dirac_oscillator} can be obtained through the substitution
$\omega\to\ -\omega$, and consists on right-handed quanta. This Hamiltonian presents similar features
as those discussed above, and can be exactly mapped onto an anti-Jaynes-Cummings interaction
\begin{equation}
H= \hbar(ga_r\sigma^-+g^*a_r^{\dagger}\sigma^+) + mc^2\sigma_z,
\end{equation}
with similar parameters. It is precisely this chirality which allows an exact mapping between the
JC, AJC, and the lefthanded and righthanded 2D Dirac oscillator. This essential property, missing in the 3D case,
 forbids an exact mapping of Eq.~\eqref{ec_dirac_oscillator_3} onto a JC-like Hamiltonian.

In conclusion, we have demonstrated the exact mapping of the 2+1 Dirac oscillator onto
 a Jaynes-Cummings model, allowing an interplay between relativistic quantum mechanics and quantum optics. We gave two relevant examples: the {\it Zitterbewegung} and collapse-revival dynamics. In addition, we showed that the implementation of a 2D Dirac oscillator in a single trapped ion, with all analogies and measured observables, is at reach with current technology.

\noindent {\acknowledgements} A.B. and M.A.MD. aknowledge DGS grant under contract
BFM2003-05316-C02-01 , and CAM-UCM grant under ref. 910758. E.S.
acknowledges finnancial support of EuroSQIP and DFG SFB 631
projects.


\end{document}